\RequirePackage{lineno}
\documentclass[twocolumn,showpacs,floatfix,amsmath,amssymb]{revtex4-2}

\usepackage{graphicx}   
\usepackage{placeins}
\usepackage{xcolor}
\usepackage{natbib}
\usepackage{hyperref}
\usepackage{booktabs}
\usepackage{amsmath,amssymb,amsfonts}
\usepackage{xspace}
\usepackage{multirow}

\newcommand{\sqrtsNN} {\ensuremath{\sqrt{s_{\mathrm{NN}}}}\xspace}
\newcommand{\sevenn}        {$\sqrt{s_{\mathrm{NN}}}~=~7$~TeV\xspace}

\newcommand{\mpt}{\ensuremath{\langle p_{\rm T} \rangle}\xspace}
\newcommand{\ppt}{\ensuremath{p_{\rm T}}\xspace}

\newcommand {\auau}  {\ensuremath{Au+Au}\xspace}
\newcommand {\pbpb}  {\ensuremath{Pb+Pb}\xspace}
\newcommand {\xexe}  {\ensuremath{Xe+Xe}\xspace}
\newcommand {\oo}  {\ensuremath{O+O}\xspace}
\newcommand {\pp}    {\ensuremath{p+p}\xspace}
\newcommand {\ppb}    {\ensuremath{p+Pb}\xspace}

\newcommand {\dndeta}       {\ensuremath{\mathrm{d}N_\mathrm{ch}/\mathrm{d}\eta}\xspace}

\newcommand {\dNdy} {\ensuremath{{\mathrm d}N/{\mathrm d}{\textsl y}}\xspace}
\newcommand {\avdndeta}     {\ensuremath{\langle\dndeta\rangle}\xspace}
\newcommand {\Nch}       {\ensuremath{{N}_\mathrm{ch}}\xspace}
\newcommand {\avnch}     {\ensuremath{\langle {N}_\mathrm{ch}\rangle}\xspace}
\newcommand{\cent}   [2] {$#1$--$#2\%$}

\usepackage[bottom]{footmisc} 

\begin{document}


\title{ Dynamics of identified particles production in oxygen-oxygen collisions at \sevenn using EPOS4}

\author{A.~M.~Khan}
 \email{ahsan.mehmood.khan@cern.ch;}
\affiliation{%
 University of Science \& Technology of China, Hefei 230026, People's Republic of China\\
}%

\author{M.~U.~Ashraf}
 \email{usman.ashraf@cern.ch; (Corresponding Author)}
\affiliation{%
Centre for Cosmology, Particle Physics and Phenomenology (CP3), Université Catholique de Louvain, B-1348 Louvain-la-Neuve, Belgium\\
}%

\author{H.~M.~Alfanda}
 \email{haidar.masud.alfanda@cern.ch;}
\affiliation{%
Key Laboratory of Quark and Lepton Physics (MOE) and Institute of Particle Physics, Central China Normal University,
Wuhan 430079, China\\
}%
\author{M.~Uzair.~Aslam}

 \email{muhammad.uzair.aslam@cern.ch;}
\affiliation{%
Pakistan Institute of Nuclear Science \& Technology, Islamabad, 44000, Pakistan\\
}%


\date{\today}

\begin{abstract}

The Large Hadron Collider (LHC) aims to inject oxygen (${}^{16}O$) ions in the next run into its experiments. This include the anticipated one-day physics run focusing on \oo collisions at center-of-mass energy \sevenn. In this study, we have used recently developed version of the EPOS model (EPOS4) to study the production of identified particles ($\pi^\pm$, $K^\pm$ and $p(\overline{p})$) in \oo collisions at \sevenn. Predictions of transverse momentum (\ppt) spectra, \mpt, integrated yield (\dNdy) for different centrality classes are studied. To provide insight into the collective nature of the produced particles, we look into the \ppt-differential particle ratios ($K/\pi$ and $p/\pi$) and \ppt-integrated particle ratios to ($\pi^++\pi^-$) as a function of charge particle multiplicity. The shape of the charge particle multiplicity (\dndeta) and mean transverse momentum (\mpt) is well described by the EPOS4. The EPOS4 predictions for the ratios of $K/\pi$ and $p/\pi$ exhibit a systematic overestimation compared to the observed trends in \pp, \ppb and \pbpb systems as a function of charged-particle multiplicity. Interestingly, the \oo results of \ppt-integrated particle ratios shows a clear final state multiplicity overlap with \pp, \ppb and \pbpb collisions. EPOS4 does not only mimics signs of collectivity, but embeds collective expansion by construction, since it relies on relativistic hydrodynamics to model the evolution of the so-called core and is one of the suitable candidates to study ultra-relativistic heavy-ion collisions. Furthermore, the foreseen data from \oo collisions at the LHC, when available, will help to better understand the heavy-ion-like behavior in small systems as well as help to put possible constraints on the model parameters.

\end{abstract}

\maketitle


\section{Introduction}
\label{sec1}

Ultra-relativistic heavy-ion collisions at the Large Hadron Collider (LHC) and the Relativistic Heavy Ion Collider (RHIC) lead to the formation of Quark-Gluon Plasma (QGP), wherein quarks and gluons are no longer confined. The experiments at both the LHC and RHIC play a crucial role in facilitating and continuously exploring the properties of this hot and dense Quantum Chromodynamics (QCD) matter~\cite{Akiba:2015jwa, BESII}. So far, several measurements have been performed in different colliding systems at different center-of-mass energies. The primary focus of these experiments is to study the properties of the QGP that flows hydrodynamically as a nearly perfect fluid~\cite{1, 2}. This is achieved by the collision of symmetric heavy-ions, such as \pbpb and \auau. However, small colliding systems, such as proton-proton (\pp), serve as a baseline for comparison.  

In 2017, LHC collided symmetric Xenon (${}^{129}Xe$) nuclei at a center-of-mass energy of {\sqrtsNN} = 5.44 TeV during a pilot run. The objective was to study a slightly smaller system size. Unexpected results were observed, including the measurement of a deformed ${}^{129}Xe$ nucleus in ultra-central collisions~\cite{3, 4, 5, 6, 7}. Apart from the deformation, most of the theoretical predictions quantitatively describe the experimental data~\cite{8, 9}. Although hydrodynamical scaling is well understood in larger systems such as \pbpb and \xexe, it is uncertain how well it holds for smaller systems~\cite{ALICE:2016kpq, ALICE:2017kwu}.

Signs of collectivity in \pp and \ppb collisions have long been reported in Refs.~\cite{Li:2012hc, ALICE:2013snk}, and the EPOS model successfully predicts the ``ridge'' structure observed in \pp collisions at the LHC energies~\cite{Werner:2010ss, Werner:2010ny}.
Recently, experiments at the LHC observed indications of the existence of deconfined matter in high-multiplicity collisions, specifically in \pp and asymmetric \ppb collisions, which is beyond the realm of heavy-ion collisions~\cite{10, 11}. The presence of QGP-like signatures in small systems at the LHC energies has drawn a lot of attention from the heavy-ion physics community. Therefore, it becomes essential to further investigate the collisions of small systems at these energies. In this context, a short run of the Oxygen-Oxygen (\oo) collision, intermediate in multiplicity between \ppb and \pbpb, is anticipated at the LHC~\cite{12, 13}. This could provide a valuable opportunity to investigate the transitions of these phenomena from large to small systems. It is important to point out that EPOS is especially adequate to study collectivity in small systems (\oo collisions here) since it had been predicting it since 2011~\cite{Werner:2010ss, Werner:2010ny, Werner:2013ipa}.   

Several recent theoretical studies have been performed to study particle production mechanisms in \oo collisions~\cite{14, 15, 16, 17, 18, 19, 20, 21, 22, Ashraf:2024ocb}. These intermediate ion collisions may provide a deeper insight to investigate the underlying mechanisms of particle production, the effects of transverse collective flow, and the production of light nuclei within the multiplicity range from small systems (\pp and \ppb) and large systems (\xexe and \pbpb)~\cite{13}. ${}^{16}O$ is a doubly magic nucleus with the distinct feature of having similar number of participating nucleons as that of \ppb system~\cite{13}, that are distributed more sparsely in the transverse plane which lead to different subsequent evolution. This unique feature of $O$ provides enhanced nuclear stability against decay and has a very compact nuclear structure~\cite{23}.   

Bulk observables, including particle spectra, charged-particle multiplicities, and particle ratios, provide an excellent probe for investigating the properties of the QGP. The correlation between these observables provide a comprehensive understanding of the interplay between soft and hard processes in the collisions, and thus shed light on the equation of state of the hot hadronic matter~\cite{24}. The recent results from the LHC revealed that final state multiplicity within an event plays a crucial role in driving the observed QGP-like properties in high multiplicity \pp collisions~\cite{10}. Therefore, it would be interesting to compare the initial and final state effects in \oo collisions, especially considering their multiplicity overlap with high-multiplicity \pp collisions. In this article, we investigate the predictions of global parameters~\cite{25} of identified particles ($\pi^\pm$, $K^\pm$ and $p(\overline{p})$) including transverse momentum \ppt spectra, charged particle multiplicity, particle ratios, and kinetic freeze-out properties in \oo collisions at \sevenn from EPOS4 simulations. We choose EPOS4 because it demonstrates good agreement with experimental data from RHIC and LHC\cite{29, 41}.

The article is organized as follows: A brief introduction to motivation is provided in Sec.~\ref{sec1}. Section~\ref{sec2} discusses the details of EPOS4. Detailed results are presented in Sec.~\ref{sec3}. Finally, the results are summarized in Sec.~\ref{sec4}.

 \section{Event Generator EPOS4}
 \label{sec2}

In this section, a brief introduction to the EPOS4 is presented.

EPOS is an Energy conserving quantum mechanical multiple scattering approach, based on Partons (parton ladders), Off-shell remnants, and Saturation of parton ladders, based on Monte carlo framework. It provides comprehensive simulations of high energy \pp and heavy-ion collisions, including both initial- and final-state dynamics. The model parameters of EPOS4 are discussed in detail in the Ref.~\cite{26, 27, 28, 29}. Various observables, such as particle production, transverse momentum \ppt distributions, and flow correlations, can be explored in both small systems (\pp collisions) and complex heavy-ion interactions. A combined approach of the Gribov-Regge Theory (GRT) with eikonalized parton model employed in the EPOS to treat the first interactions just after the collisions. In this approach, the conservation laws are satisfied and the subsequent pomerons (interactions) are treated equally~\cite{30}.  
The formalism to calculate the particle production is based on the Feynman diagrams of the QCD-inspired effective field theory, providing the energy conservation, and is the same as used for the cross-section calculations. The nucleons are considered to be composed of ``constituents'' carrying certain fraction of the incident momentum of the nucleon. The sum of these fractions equals unity ensuring that the momentum is conserved during the collisions. A nucleon termed as a ``spectator'' does not participate directly in the interaction. If a nucleon is directly participating the interaction, it can either be a ``participant'' or it can be a ``remnant''. The particle production in EPOS does not rely on the so-called ``Lund string model'' but on a slightly different string model (quite similar yet different on some technical aspects)~\cite{Werner:1993uh}.

In high multiplicity \pp and heavy-ion collisions, the density of the ``strings'' during particle interactions can become very high and the individual strings are unable to decay independently. EPOS addresses this issue by introducing a dynamic process of division of the segments of the strings into the ``core'' and the ``corona''~\cite{32, 32a, 32b}.  Recently, the author has introduced a new developments in the EPOS framework, referred as EPOS4~\cite{26, 27, 28, 29, Werner:2023mod}. In EPOS4, a new understanding of strong interaction is developed between four major concepts used in \pp and $AA$ collisions. These concepts include rigorous scattering, energy conservation, factorization, and saturation~\cite{Werner:2023mod}. This formalism effectively measures high \ppt particle production in factorization mode, and simulates the collective effects in high-multiplicity events. Importantly, the implementation of a dynamical saturation scale in this version does not impact high \ppt particle production, even if a large number of parallel scatterings occurs~\cite{26}. Further details about the EPOS4 can be found in Refs.~\cite{26, 27, 28, 29, Werner:2023mod}. 

We produce $\approx$ 1.5 Million minimum bias (all collisions with at least one inelastic interaction) events for \oo collisions at \sevenn from EPOS4.

We generated EPOS4 simulations by using these specific parameters to investigate the behavior of \oo system. To account for core-corona effects, we set the ``core'' option to``full''. Additionally, the ``hydro'' parameter was enabled to incorporate the effects of hydrodynamic evolution of the system, which describes the collective flow of particles. The equation of state (EoS), a crucial component defining the system's behavior under pressure and density changes, was left as the standard option offered by EPOS4. Furthermore, the hadronic cascade, the interactions between particles after the initial collision, was turned on. It is important to note that the identified particle spectra obtained from these simulations have not been corrected for contributions from weak decays of short-lived particles. The centrality classes have been determined by the reference multiplicity ($Refmult$) observable, which measures the total number of particles produced in an event at $|\eta| < 0.5$. The centrality percentile has been assigned on the basis of the $Refmult$ distribution, i.e. from central to peripheral collisions.

\section{Discussion}
\label{sec3}

In this section, we present the predictions for different global properties of identified particles ($\pi^\pm$, $K^\pm$ and $p(\overline{p})$), such as charged particle multiplicity (\dndeta), transverse momentum {\ppt} spectra, integrated yield (\dNdy) and \ppt-differential and \ppt-integrated particle ratios ($K/\pi$ and $p/\pi$) for different centrality classes in \oo collisions at \sevenn using EPOS4. From here onwards, $(\pi^+ + \pi^-)$, $(K^{+} + K^-)$ and  $(p +\overline{p})$ are denoted as pions ($\pi$), kaons ($K$) and protons ($p$) respectively. 

\subsection{Charged particle multiplicity distributions}

Charged particle pseudorapidity density distributions, \avdndeta is an important observable in heavy-ion collisions. The pseudorapidity, $\eta$, is defined as $\eta = - \rm{ln} [\rm {tan}(\theta/2)]$, where $\theta$ is the emission angle relative to the beam direction. \avdndeta is proportional to the entropy density at freeze-out. As the entropy density of a closed system tends to increase with collision energy~\cite{39}, pseudorapidity density provide insights into the initial-state density of partons and any further entropy produced during subsequent evolution~\cite{39}. Additionally, soft processes are sensitive to the \avdndeta in the collision, while the mean transverse mass and momentum provides insight into the hard processes. 

\begin{figure}[thb]
\includegraphics[width=0.45\textwidth]{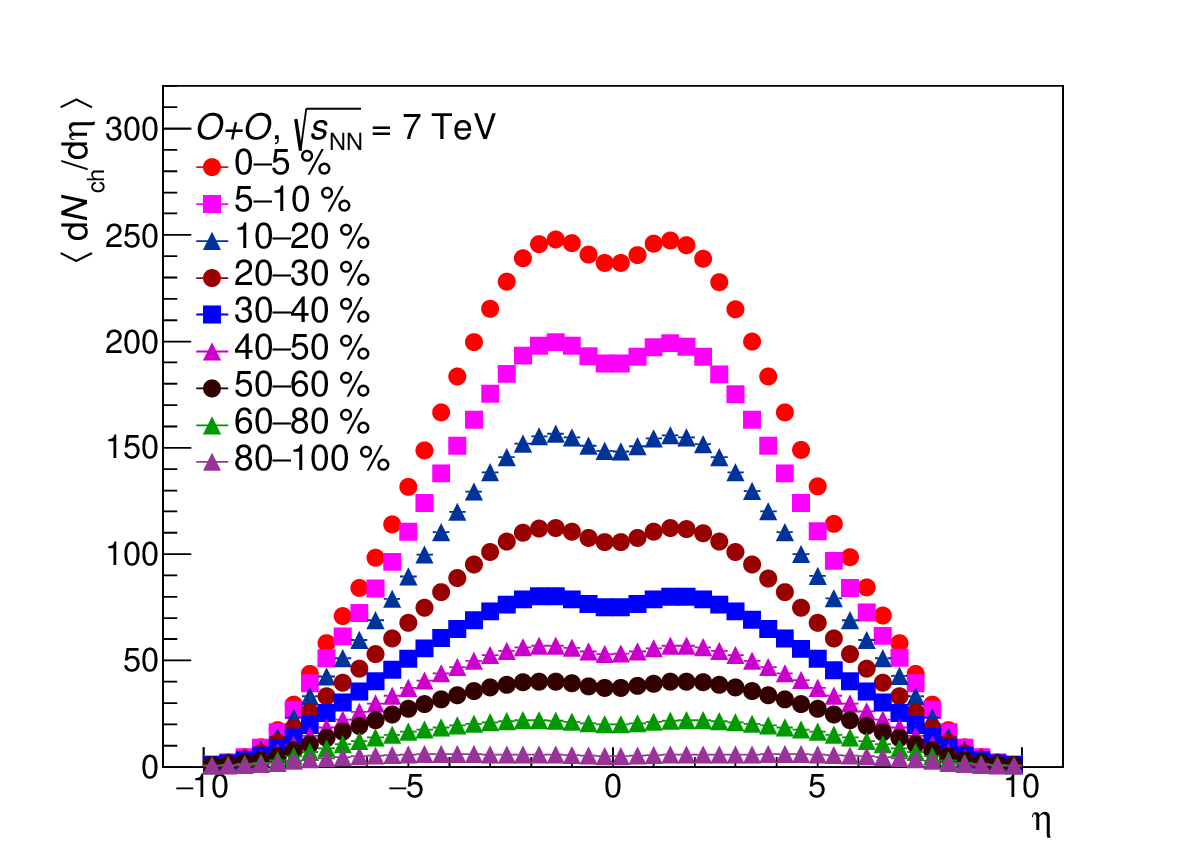}
\caption{(Color online) Pseudorapidity distributions of charged particles in \oo collisions at \sevenn for different centrality classes using EPOS4.}
\label{fig1}
\end{figure}

\begin{table}[htbp]
  \centering
  \caption{Average charged-particle multiplicity at $|\eta| < 0.5$ in \oo collisions at \sevenn for different centrality classes using EPOS4.}
    \begin{tabular}{cccc}
    \toprule
    Centrality ($\%$) & \avdndeta & $\langle{N_{\rm part}}\rangle\pm{\rm rms}$ & $\langle{N_{\rm coll}}\rangle\pm{\rm rms}$ \\
    \midrule
    \midrule
            $0$~--~$5$ & $236.44\pm0.14$ & $27.86\pm2.18$  &$49.89\pm9.32$ \\
            $5$~--~$10$ &$189.801\pm0.13$ &$25.05\pm2.33$ &$40.77\pm8.79$\\ 
            $10$~--~$20$ &$148.437\pm0.08$ &$21.44\pm3.15$ &$31.35\pm7.88$\\ 
            $20$~--~$30$ &$105.863\pm0.07$ &1$6.87\pm3.12$&$20.84\pm5.65$\\ 
            $30$~--~$40$ &$75.027\pm0.06$ &$12.44\pm2.56$ &$13.35\pm4.11$\\ 
            $40$~--~$50$ &$53.037\pm0.05$ &$9.60\pm2.52$ &$9.16\pm3.24$\\ 
            $50$~--~$60$ &$37.146\pm0.06$ &$6.98\pm2.3$ &$5.94\pm2.54$\\ 
            $60$~--~$80$ &$19.889\pm0.02$ &$4.49\pm1.96$ &$3.31\pm1.84$\\ 
            $80$~--~$100$ &$5.127\pm0.01$ & $1.89\pm1.53$&$1.13\pm1.06$\\ 
    \bottomrule
    \bottomrule
    \end{tabular}
  \label{tab1}
\end{table}

Fig.~\ref{fig1} shows pseudorapidity distributions of the charged particles from most central (\cent{0}{5}) to most peripheral (\cent{80}{100}) \oo collisions at \sevenn using EPOS4. It is observed that the EPOS4 is capable of describing the typical shape of the \avdndeta distribution and is consistent with the recent AMPT studies in \oo collisions at \sevenn reported in Ref.~\cite{22}. It has been reported in Ref.~\cite{Werner:2023mod} that \avdndeta distribution in \auau at 200 GeV as well as particle production in at 5.02 TeV is pretty well described by EPOS4. The values of \avdndeta for different centrality classes are listed in Tab.~\ref{tab1}. The extracted values of $\langle \Nch \rangle$ and the \mpt from \oo collisions at \sevenn in the kinematic range $|\eta| < 0.8 $ and 0.15 GeV/$c <$ \ppt $<$ 10 GeV/$c$ are listed in Tab.~\ref{tab2}. We only considered the events with $\Nch > 0$ to be consistent with the already reported~\cite{ALICE:2022xip} results from different systems. It is observed that $\langle \Nch \rangle$ values from \oo collisions at \sevenn lies between \ppb and \xexe collisions.

\begin{table}[htbp]
    \centering
    \caption{ $\langle \Nch \rangle$ and the \mpt extracted from \oo collisions at \sevenn in the kinematic range $|\eta| < 0.8 $ and 0.15 GeV/$c <$ \ppt $<$ 10 GeV/$c$ and only events with $\Nch > 0$  are considered. The corresponding $\langle \Nch \rangle$ values from \pp, \ppb, \xexe and \pbpb collisions are taken from Ref.~\cite{ALICE:2022xip} while $\langle \Nch \rangle$ values from \oo collisions at \sevenn are from present work.}
    \begin{tabular}{cccc}
        \toprule
              & \sqrtsNN (TeV) & \avnch         & \mpt (MeV/$c$) \\
       \midrule
        \midrule
      {      \multirow{5}[1]{*}  \pp}   & $2.76$         & $7.18\pm0.24$  & $589.7\pm2.6$  \\
         & $5.02$         & $8.21\pm0.10$  & $612.2\pm2.7$  \\
         & $7$            & $8.86\pm0.12$  & $627.1\pm1.6$  \\
          & $8$            & $9.05\pm0.22$  & $631\pm5$      \\
          & $13$           & $10.31\pm0.09$ & $654\pm1.0$    \\
           \hline
       {      \multirow{2}[0]{*}  \ppb} & $5.02$         & $25.51\pm0.25$ & $711.9\pm1.3$  \\
          & $8.16$         & $29.56\pm0.26$ & $741.5\pm1.4$  \\
           \hline
        \oo   & $7$            & $86.8\pm1.2$  & $679.8\pm1.9$  \\
        \xexe & $5.44$         & $458\pm10$     & $717.4\pm1.8$  \\
         \hline
        {      \multirow{2}[0]{*} \pbpb } & $2.76$         & $573\pm9$      & $687.3\pm1.3$  \\
         & $5.02$         & $682\pm13$     & $724.1\pm1.1$  \\
        \bottomrule
        \bottomrule
    \end{tabular}
    \label{tab2}
\end{table}

\subsection{Transverse momentum (\ppt) spectra}

In ultra-relativistic heavy-ion collisions, the observable, such as production yield and/or \ppt spectra, are important to explore the particle production mechanisms. Additionally, the \ppt spectra of identified particles in heavy-ion collisions provide deeper insights on the transverse expansion of the QGP and the freeze-out properties of the hadronic phase~\cite{Andronic:2020iyg, Andronic:2017pug, Andronic:2014zha, Tariq:2024hfc}. 

\begin{figure}[thb]
\includegraphics[width=0.48\textwidth]{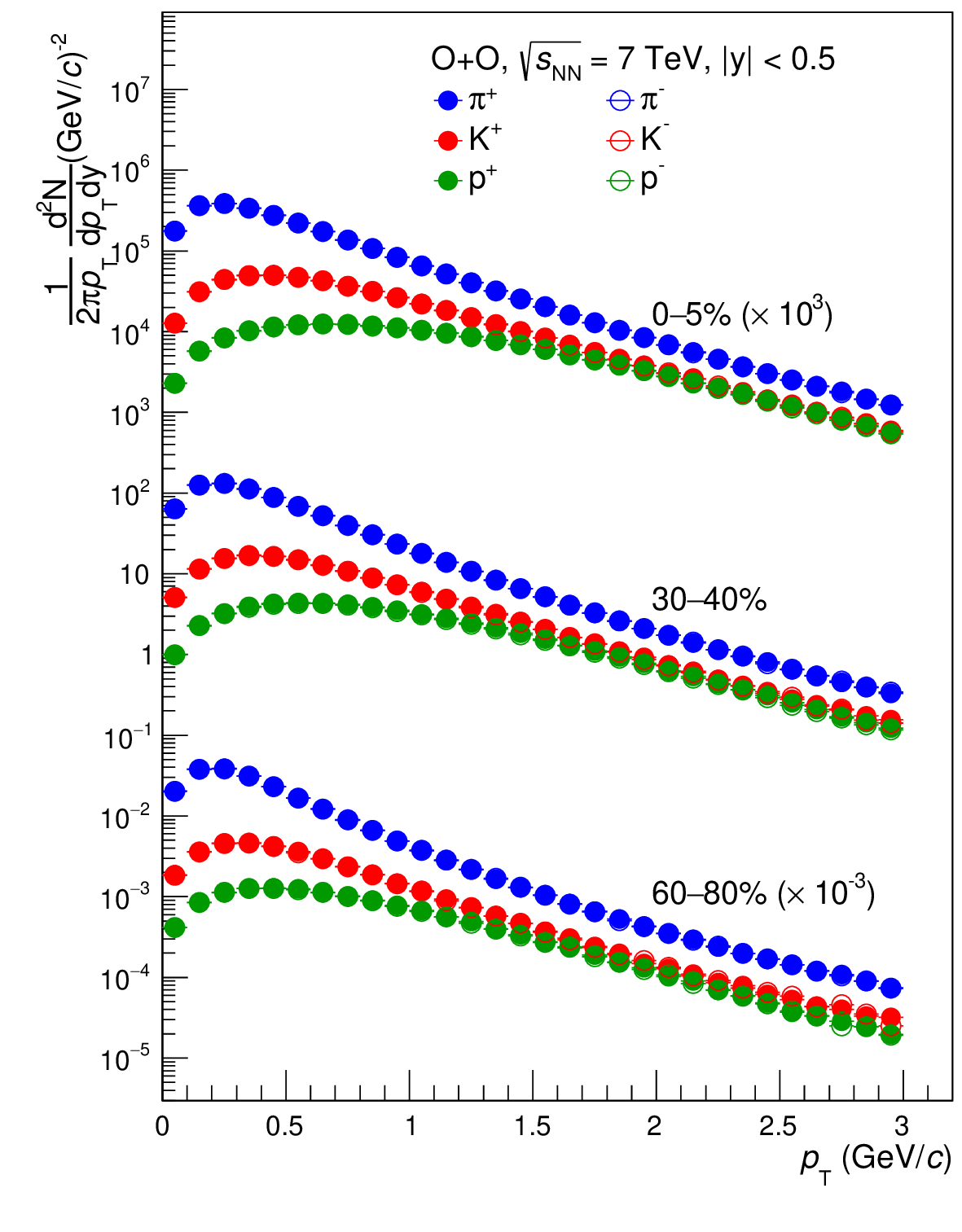}
\caption{(Color online) Transverse momentum \ppt spectra of identified particles at mid-rapidity in central (\cent{0}{5}),  mid-central (\cent{30}{40}) and peripheral (\cent{60}{80}) \oo collisions at \sevenn using EPOS4. Different color show different particle species. The spectra are scaled by different powers of 10 for better visualization.}
\label{fig2}
\end{figure}

The {\ppt} spectra of the identified particles in central (\cent{0}{5}), mid-central (\cent{30}{40}) and peripheral (\cent{60}{80}) \oo collisions at \sevenn using EPOS4 is shown in Fig.~\ref{fig2}. Solid symbols are for positive particles whereas open symbols are for negative particle species. Pions are the lightest hadrons and consequently the most abundantly produced particle. A clear mass-dependent behavior of the produced particles is observed at low-{\ppt} which is similar to the results already reported from \pbpb collisions at \sqrtsNN = 2.76 TeV from the ALICE experiment at LHC~\cite{41a}. The spectra of all particle species shows a dip at {\ppt} $< 0.5$ GeV/$c$ and kaons and protons spectra approach the pion spectra at intermediate-{\ppt}. This observation can be attributed to the effects of radial flow within the medium as the particles are pushed from low-{\ppt} to intermediate-{\ppt} due to radial flow~\cite{41b}. The observed shapes of the {\ppt}-spectra can potentially be explained by two mechanisms. At low and intermediate {\ppt}, the coalescence mechanism might play a dominant role~\cite{41c}. This mechanism proposes that particles are formed by the combination of pre-existing hadrons. Alternatively, the production of high-{\ppt} jets through the fragmentation process could also contribute to the spectra~\cite{41d, 41e}. However, the effects of fragmentation are likely more significant beyond the intermediate-{\ppt} region. EPOS4 successfully reproduces the shape of the {\ppt} spectra of identified particles. Recent reports (Refs.~\cite{29, 41}) indicate that EPOS4 provides a good description of the experimental data at the RHIC and LHC energies.

\subsection{Integrated yield (\dNdy)}
In this section, we show the integrated yields (\dNdy) for each particle species as a function of centrality in \oo collisions at \sevenn.

\begin{figure}[!ht]
\includegraphics[width=0.48\textwidth]{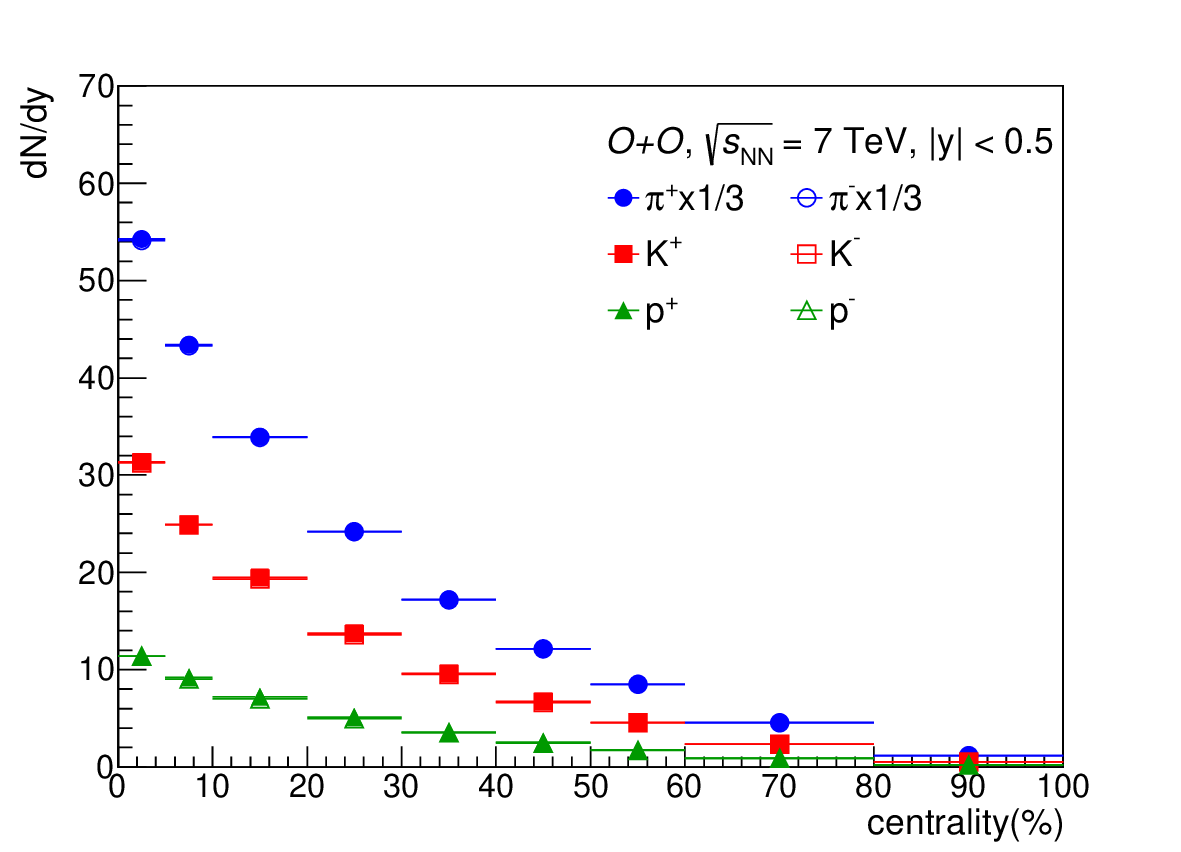}
\caption{(Color online)  Integrated yield, \dNdy of $\pi^\pm$, $K^\pm$ and $p(\overline{p})$ at mid-rapidity as a function of centrality in \oo collisions at \sevenn using EPOS4. Different symbols show different particle species.}
\label{fig5}
\end{figure}

Figure~\ref{fig5} shows the integrated yield (\dNdy) of pions, kaons, and protons as a function of centrality in \oo collisions at \sevenn. Different symbols show different particle species. The \dNdy of pions is scaled for better visualization. As expected, pions are the most abundantly produced particles among those identified. This leads to a higher integrated yield for pions compared to kaons and protons. This observation aligns with the predictions of thermalized Boltzmann production of secondary particles in high-energy nuclear collisions.

\subsection{Mean Transverse momentum (\mpt)}

In this section, we present the mean transverse momentum (\mpt) of pions, kaons and protons as a function of collision centrality in \oo collisions at \sevenn using EPOS4.

\begin{figure}[thb]
\includegraphics[width=0.48\textwidth]{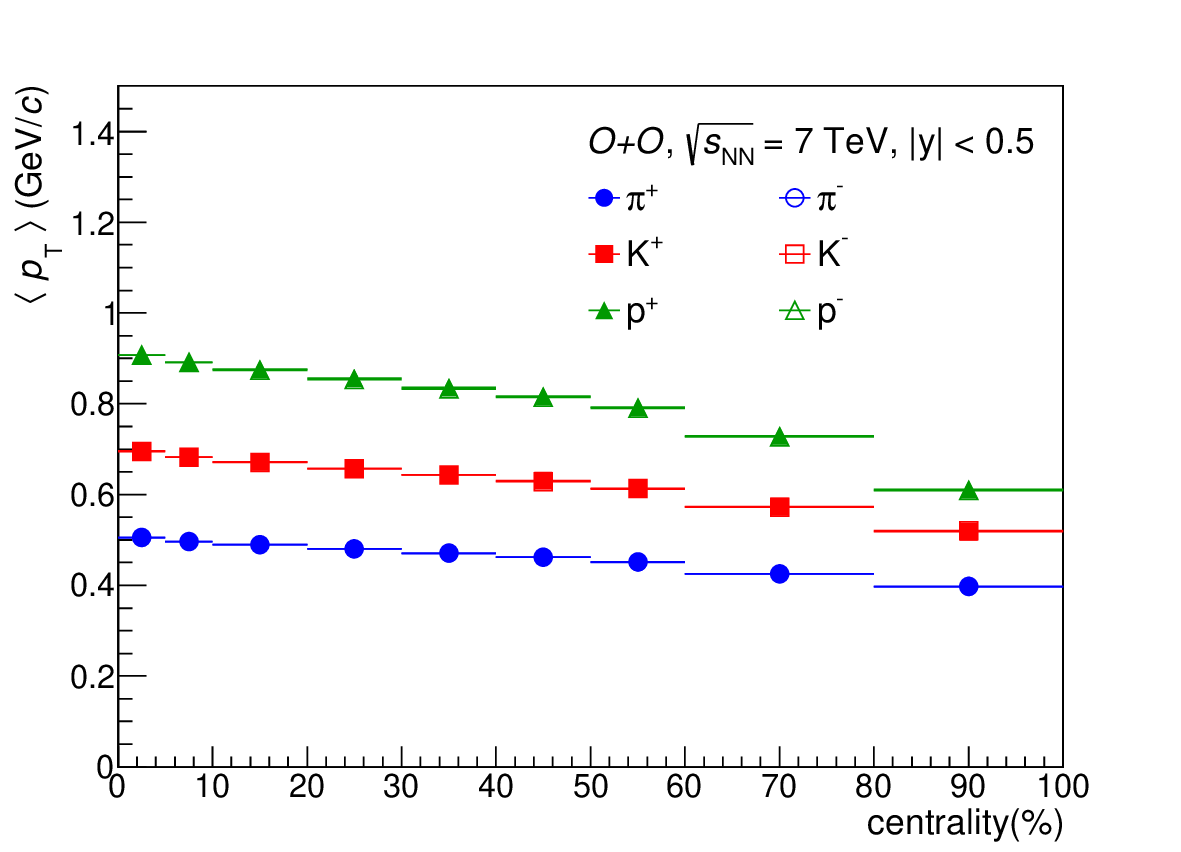}
\caption{(Color online)  \mpt of pions, kaons and protons at mid-rapidity as a function of centrality in \oo collisions at \sevenn using EPOS4. Different symbols show different particle species. }
\label{fig6}
\end{figure}

Figure~\ref{fig6} displays the centrality dependence of \mpt in \oo collisions at \sevenn, as predicted by the EPOS4. The EPOS4 simulations show an increasing trend in \mpt from peripheral to central collisions. This suggests a corresponding increase in radial flow in more central collisions. The \mpt is related to the shape (slope) of the particle momentum spectra. A similar trend of increasing \mpt with centrality has been observed in \pbpb collisions at a center-of-mass energy of {\sqrtsNN}= 5.02 TeV. The EPOS4 successfully reproduces this trend observed in the experimental data for these \pbpb collisions~\cite{29}.

\begin{figure}[thb]
\includegraphics[width=0.48\textwidth]{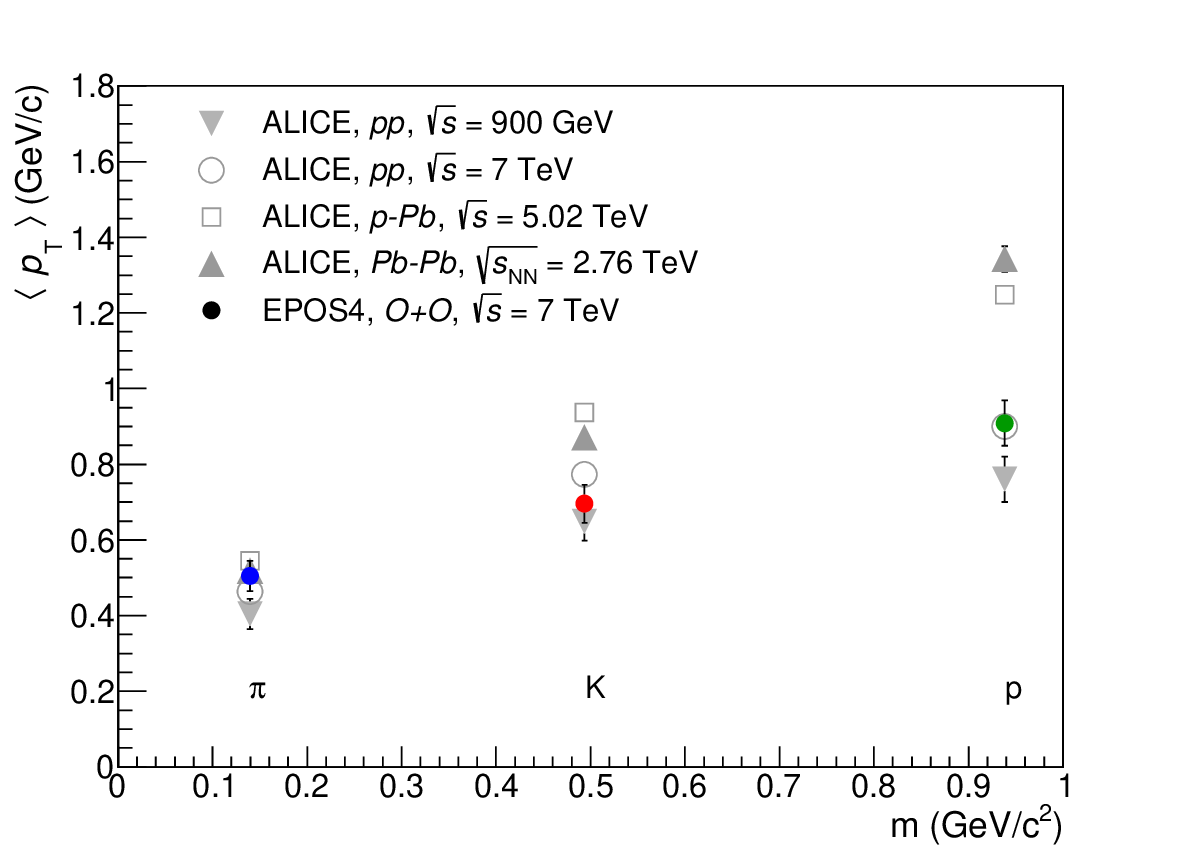}
\caption{(Color online) \mpt pions, kaons and protons at mid-rapidity as a function of particle mass in \oo collisions at \sevenn using EPOS4. The results are compared with \pp collisions at {\sqrtsNN}= 900 GeV~\cite{42} and 7 TeV~\cite{39a}, \ppb collisions at \sqrtsNN = 5.02 TeV~\cite{42a} and \pbpb collisions at \sqrtsNN = 2.76 TeV~\cite{41a}. Colored symbols represent the predictions from EPOS4 from \oo collisions at \sevenn, while open gray symbols represent the published data from different collision systems at different energies.}
\label{fig7}
\end{figure}

Figure~\ref{fig7} compares the \mpt of identified particles as a function of particle mass in \oo collisions at \sevenn with experimental data from $pp$ collisions at \sqrtsNN = 900 GeV~\cite{42} and 7 TeV\cite{39a}, \ppb collisions at \sqrtsNN = 5.02 TeV~\cite{42a} and \pbpb collisions at \sqrtsNN = 2.76 TeV~\cite{41a}. EPOS4 simulations show a slight increase in the \mpt with increasing center-of-mass energy. This observation is interesting because the momentum spectra themselves suggest an increasing contribution from hard scattering processes, which typically lead to higher \mpt particles. Additionally, EPOS4 simulations for \oo collisions at \sevenn show a clear ordering of \mpt with particle mass. This aligns with the trend observed in previous studies~\cite{42, 43, 44, 45}.

\subsection{Particle ratios}
In this section, we present the predictions of \ppt-differential ratios of pions, kaons and protons in \oo collisions at \sevenn using EPOS4.

The particle production mechanisms depend heavily on the range of \ppt. For example, at intermediate \ppt, the dominant process is thought to be coalescence. In contrast, fragmentation takes over as the main mechanism at high \ppt. Because of this \ppt dependence, studying the ratios of different particle types as a function of transverse momentum (\ppt-differential particle ratios) is a valuable approach. This investigation of \ppt-differential particle ratios is one of the main focus of the present work.

\begin{figure}[thb]
\includegraphics[width=0.48\textwidth]{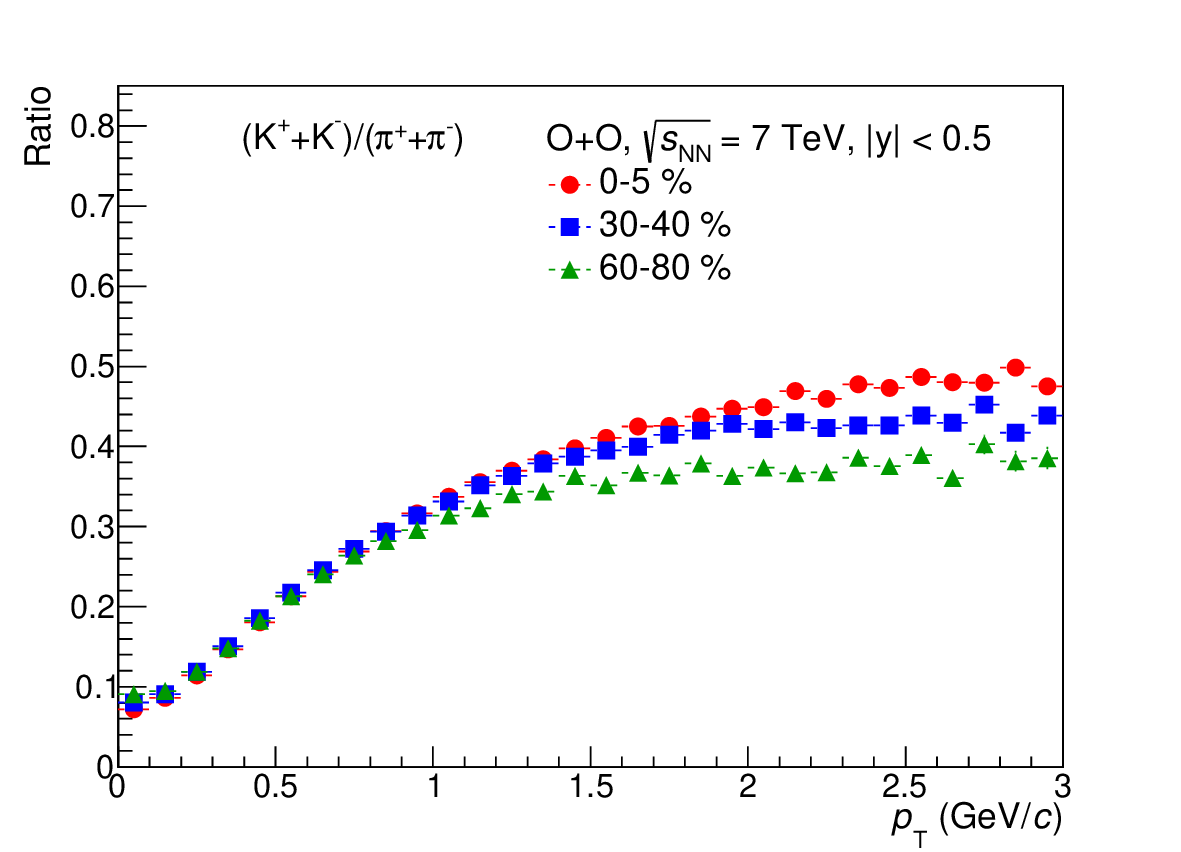}
\caption{(Color online) \ppt-differential kaon-to-pion ratio at mid-rapidity in \oo collisions at \sevenn for central (\cent{0}{5}), mid-central (\cent{30}{40}) and peripheral (\cent{60}{80}) centrality classes using EPOS4. Different symbols show various centrality bins.}
\label{fig8}
\end{figure}

Figure~\ref{fig8} presents the \ppt-differential kaon-to-pion ratio in central (0--5\%), mid-central (30--40\%), and peripheral (60--80\%) \oo collisions at \sevenn, as predicted by the EPOS4. This ratio is considered a measure of strangeness enhancement, where strange particles (kaons in this case) are produced more frequently relative to pions. EPOS4 simulations show an enhancement of strangeness production with increasing \ppt in \oo collisions at \sevenn. At low-\ppt, this enhancement exhibits a weak dependence on centrality, with similar ratios observed in both central and peripheral collisions. However, at intermediate-\ppt, a strong centrality dependence is observed. Here, strangeness production is highest (up to $0.5$) in central (0--5\%) collisions and relatively lower (up to $0.35$) in peripheral (60--80\%) collisions. The observed centrality dependence of particle ratios is also seen in \ppb~\cite{42a} and \pbpb~\cite{41a} collisions.

\begin{figure}[thb]
\includegraphics[width=0.48\textwidth]{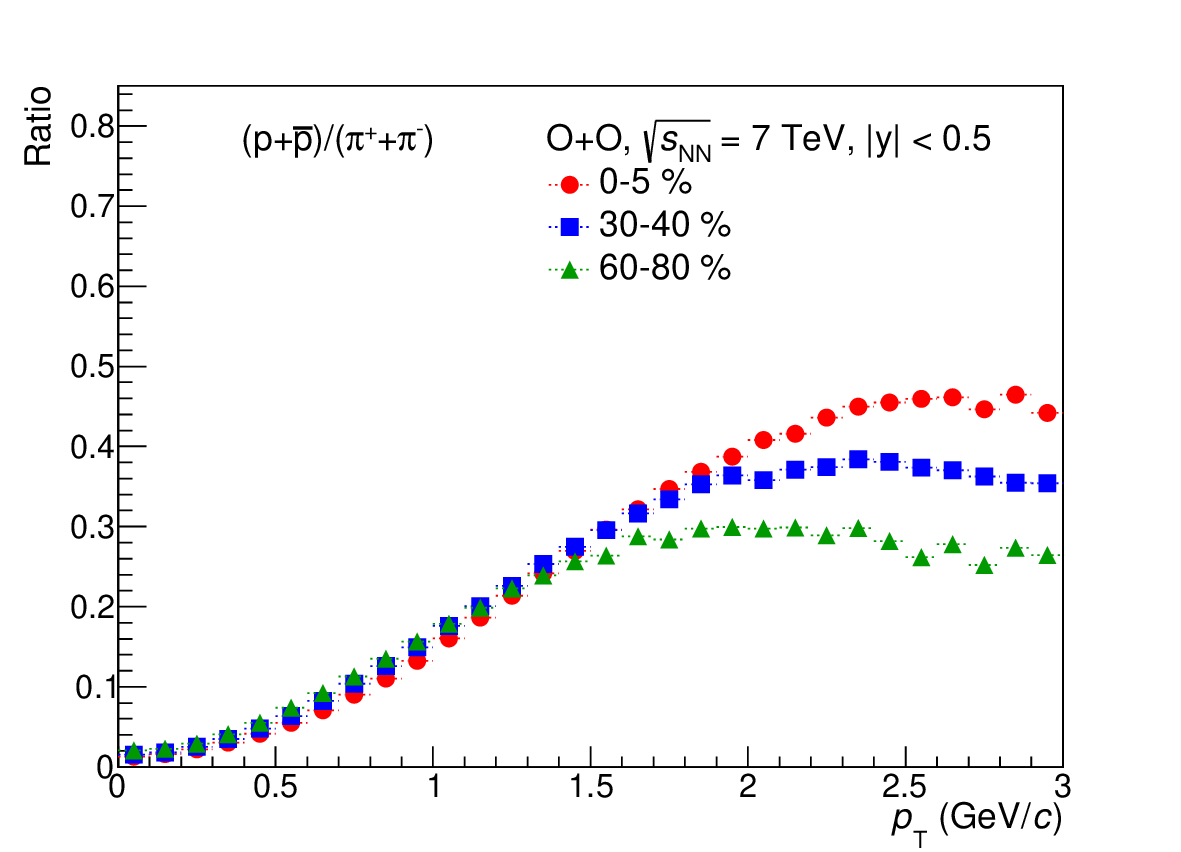}
\caption{(Color online) \ppt-differential pion-to-proton ratio in \oo collisions at \sevenn for central ( \cent{0}{5}), mid-central (\cent{30}{40}) and peripheral (\cent{60}{80}) centrality classes using EPOS4. Different symbols show various centrality bins.}
\label{fig9}
\end{figure}

Figure~\ref{fig9} displays the \ppt-differential proton-to-pion ratio (lightest baryon to lightest meson) in \oo collisions at \sevenn, as predicted by the EPOS4. This ratio serves as an indicator of the relative production of baryons compared to mesons. The EPOS4 simulations show an increasing trend in the ratio (up to a maximum of $0.45$) at intermediate \ppt for (0--5\%) central collisions. In contrast, the trend appears to plateau at intermediate \ppt in mid-central (30--40\%) and peripheral (60--80\%) collisions, similar to the trend observed in \pp, \ppb and \pbpb collisions at the LHC~\cite{39a, 42a, 41a}. Overall, the EPOS4 successfully reproduce the general shapes of both ratios in \oo collisions at \sevenn compared to experimental data from other systems~\cite{39a, 42a, 41a}.

\begin{figure}[thb]
\includegraphics[width=0.48\textwidth]{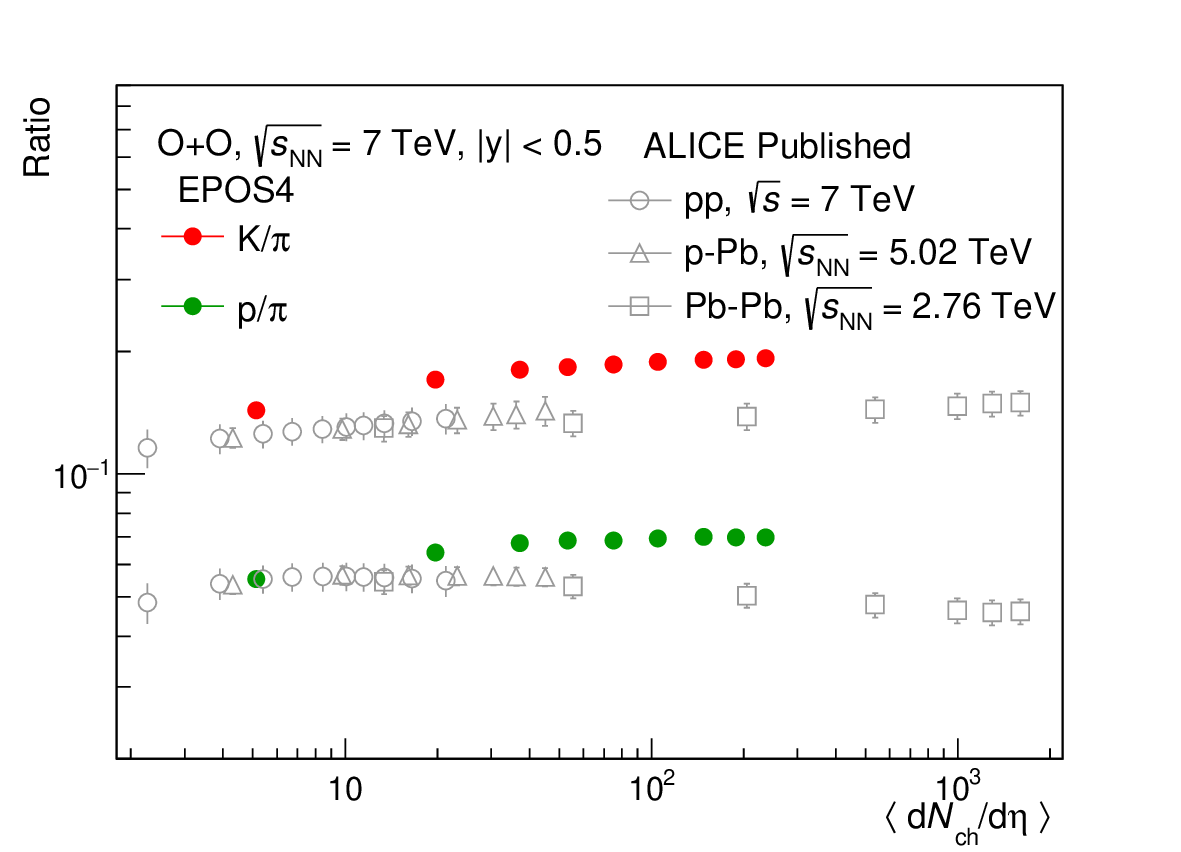}
\caption{(Color online) Multiplicity dependent \ppt-integrated proton-to-pion and kaon-to-pion ratio in \oo collisions using EPOS4. Colored symbols represent the predictions from EPOS4 from \oo collisions at \sevenn, while open gray symbols represent the published data from different collision systems at different energies.}
\label{fig10}
\end{figure}

It is very interesting to note that from EPOS4 simulations, the \ppt-differential particle ratios (\ppt $> 1$ GeV/$c$) exhibit a strong centrality dependence, while the \ppt-integrated particle ratios shows very small dependence on centrality. This suggests that the relative production of different particle types with respect to pions does not strongly depend on centrality. This is due to the fact that the \ppt-integrated yield is mostly influenced by the low-\ppt particles (\ppt $< 1$ GeV/$c$). Figure~\ref{fig10} shows the \ppt-integrated proton-to-pion and kaon-to-pion ratio as a function of charge particle multiplicity for \oo collisions at \sevenn using EPOS4. These ratios are compared to the data for \pp collisions at $\sqrt{s}$ = 7 TeV, \ppb collisions at \sqrtsNN = 5.02 TeV and \pbpb collisions at \sqrtsNN = 2.76 TeV from ALICE experiment~\cite{39a, 42a, 41a}. 

Considering charged-particle multiplicity as proxy for the size of the colliding system, the EPOS4 predicts a trend in the \ppt-integrated proton-to-pion ratio for \oo collisions at \sevenn that is opposite to observations in \pp, \ppb and \pbpb collisions~\cite{39a, 42a, 41a}. The \ppt-integrated proton-to-pion ratio in \oo collisions predicted by EPOS4 exhibits an increasing trend with increasing charged-particle multiplicity, whereas data from ALICE shows a decreasing trend with increasing multiplicity.

However, the EPOS4 predictions for the \oo system slightly overpredict the measured proton-to-pion ratios in both smaller and larger collision systems. It is worth noting that similar overestimations are observed for the proton-to-pion ratios predicted by EPOS4 in \pbpb collisions at \sqrtsNN = 2.76 TeV compared to experimental results~\cite{29}. Figure~\ref{fig10} also compares the EPOS4 predictions for the \ppt-integrated kaon-to-pion ($K/\pi$) ratio with the experimental data from \pp collisions at $\sqrt{s}$ = 7 TeV, \ppb collisions at \sqrtsNN = 5.02 TeV and \pbpb collisions at \sqrtsNN = 2.76 TeV from ALICE experiment~\cite{39a, 42a, 41a}. EPOS4 simulations for \oo collisions at \sevenn suggest a slight increase in the $K/\pi$ ratio with increasing centrality. This trend aligns with observations in \pp, \ppb and \pbpb collisions, suggesting a possible system-size dependence of strangeness enhancement. However, the EPOS4 overestimates the magnitude of this effect compared to existing experimental data~\cite{39a, 42a, 41a}.

An intriguing observation is that the EPOS4 predicts a clear final state multiplicity overlap for both the proton-to-pion and kaon-to-pion ratios in \oo collisions with \pp, \ppb and \pbpb collisions. Similar observations for the ratios of strange baryons relative to pions has recently been reported in Ref.~\cite{Ashraf:2024ocb}. EPOS4 results for \oo collisions are clearly above all experimental results from other systems. It is important to note that this experimental data for both ratios comes from different collision systems and energies. Future experimental data from \oo collisions will be essential for a more conclusive understanding of this observation and the ability of the EPOS4 to accurately predict strangeness production in small systems.

\section{Conclusions}
\label{sec4}

This study investigates the production of various identified particles ($\pi^\pm$, $K^\pm$ and $p(\overline{p})$) in \oo collisions at \sevenn using recently developed EPOS4. The key findings of this investigation are summarized below:

\begin{itemize}
    \item The EPOS4 successfully reproduces trend in the shape of all charged-particle multiplicity (\dndeta) distribution as compared to other systems. This prediction can be compared with future experimental data from \oo collisions at \sevenn. Such a comparison will be crucial for validating the model's capability of describing particle production in this specific collision system.
    
    \item EPOS4 successfully reproduces the shapes of the \ppt spectra for various identified particles. Notably, the \ppt spectra exhibit a mass dependence at low-\ppt, consistent with observations in \pbpb collisions at \sqrtsNN = 2.76 TeV. The convergence of heavier particle spectra with pions at intermediate-\ppt provides evidence for the presence of radial flow in \oo collisions.
    
    \item \ppt-integrated yield (\dNdy) normalized by total events of identified particles exhibits a clear mass ordering. The production of heavier particle is lower and decreases towards peripheral collisions. The dominance of pions production is a well-established phenomenon attributed to their lower mass, making them easier to produce in these collisions.

    \item We observe increase in average transverse momentum (\mpt) with increasing collision centrality indicating stronger radial flow in more central collisions. The EPOS4 successfully reproduces this mass ordering of \mpt observed in \oo collisions, follows the trend established by the published results from various collision systems. 

    \item It is observed that the \ppt-differential $K/\pi$ ratio shows increasing trend with increase in \ppt and saturate towards higher \ppt. In contrast, \ppt-differential $p/\pi$ ratio shows a plateau around 3 GeV/$c$ which might be indicative of radial flow like patterns.

    \item Our analysis reveals an increase in strangeness enhancement with increasing \ppt. This effect is more pronounced in central collisions for particles at intermediate-\ppt. At low-\ppt this effect becomes insignificant. 

    \item It is interesting to note that the multiplicity dependence of \ppt-integrated $K/\pi$ ratio shows increasing trend with centrality similar to \pp, \ppb and \pbpb collisions suggesting possible system-size dependence of strangeness enhancement. On the other hand, \ppt-integrated $p/\pi$ ratio from EPOS4 predicts an opposite trend compared to what observed in \pp, \ppb and \pbpb collisions. The EPOS4 predictions for both ratios in \oo collisions at \sevenn are slightly overestimates the existing experimental data from different systems. It would be interesting to compare these observations with the foreseen experimental data, when available for more conclusive understanding of this observation. Additionally, EPOS4 also predicts a final state multiplicity overlap in between \oo collisions and those observed in \pp, \ppb and \pbpb collisions.

\end{itemize}

The foreseen data from \oo collisions at the LHC, when available, will help to better understand the heavy-ion-like behavior in small systems and help to put possible constraints on the model parameters.

\section{Acknowledgements}

The authors are grateful to K. Werner for valuable discussions regarding recently developed EPOS4 framework and to M. Aamir Shahzad and Jagbir Singh for insightful discussions and comments on the manuscript. We would also like to thank Giorgio Mauceri for his careful review of the manuscript and his constructive suggestions.

\bibliographystyle{utphys}
\bibliography{bib}

\end{document}